\documentclass[a4paper,11pt]{article}
\usepackage[left=2cm, right=1.5cm, top=2cm, bottom=2cm, includehead=false, includefoot=false]{geometry} 
\usepackage{float}
\usepackage{amssymb}

\usepackage{graphicx}% remove demo option in actual document

\usepackage{fancyhdr}
\usepackage{authblk}

\title{\textbf{State-of-the-Art on Query \& Transaction Processing Acceleration}}
\author[1]{Bernd Amann}
\author[1]{Youry Khmelevsky}
\author[2]{Ga\'etan Hains}
\affil[1]{LIP6, Campus Pierre et Marie Curie, Sorbonne Universit\'e, Paris}
\affil[2]{Huawei Technologies, Paris Research Center}

\date{\bigskip \bigskip \textbf{HUAWEI Technical Report CSI-PARIS-TR-2018-10}\\ \bigskip 2018--09--27 \\\bigskip\bigskip\bigskip\bigskip\bigskip 
{Huawei Technologies\\
2012Labs/CSI/DPSL/CSI-PARIS}}

\begin{document}
\clearpage\maketitle
\thispagestyle{empty}

\begin{figure}[b]
\hfill%
\begin{minipage}{0.6\textwidth}\raggedright
\textbf{Huawei Paris R\&D Center}
\end{minipage}
\noindent
\begin{minipage}{0.3\textwidth}% adapt widths of minipages to your needs
\includegraphics[width=2 cm]{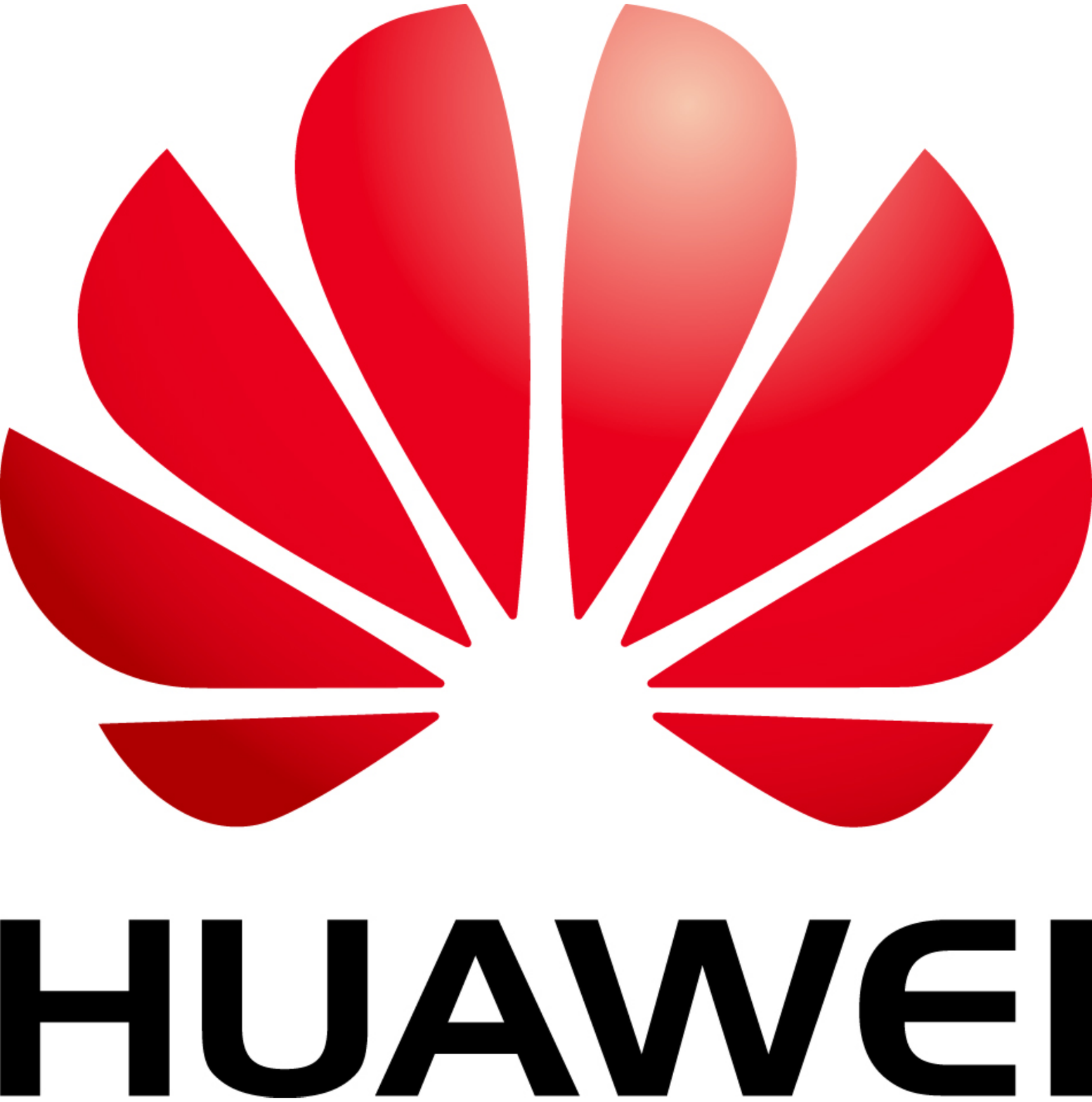}
\end{minipage}%\\

\bigskip\bigskip

\center\textbf{Technical Report CSI-PARIS-TR-2018-10	             Huawei Unclassified}
\end{figure}

\newpage

\abstract{The vast amount of processing power and memory bandwidth provided by modern Graphics Processing Units (GPUs) make them a  platform for data-intensive applications. The database community identified GPUs as effective co-processors for data processing. In the past years, there were many approaches to make use of GPUs at different levels of a database system.

In this Internal Technical Report, based on the \cite{Bre2014} and some other research papers, we identify possible research areas at LIP6 for GPU-accelerated database management systems. We describe some key properties, typical challenges of GPU-aware database architectures, and identify major open challenges.

\section{Introduction \cite{Bre2014}}

Modern processors are constrained to a certain amount of power they may consume (i.e., the power wall \cite{Borkar:2011:FM:1941487.1941507}) and further increasing frequency and parallelism would make them overly power hungry. Therefore, hardware vendors are forced to create processors that are optimized for a certain application field. These developments result in a highly heterogeneous hardware landscape, which is expected to become even more diverse in the future \cite{Borkar:2011:FM:1941487.1941507}. In order to keep up with the performance requirements of the modern information society, tomorrow's database systems will need to exploit and embrace this increased heterogeneity.

The GPU is the pioneer of modern co-processors, and --- in the last decade --- it matured from a highly specialized processing device to a fully programmable, powerful co-processor. This development inspired the database research community to investigate methods for accelerating database systems via GPU co-processing. Several research papers and performance studies demonstrate the potential of this approach \cite{Ailamaki:2001:WRC:645927.672367}, \cite{Bakkum:2010:ASD:1735688.1735706}, \cite{He:2009:RQC:1620585.1620588}, \cite{He:2013:RCH:2536206.2536216}, \cite{6816677} --- and the technology has also found its way into commercial products (e.g., Jedox \cite{PaloGPUaccelerator} or ParStream \cite{Parstream}).

Using graphics cards to accelerate data processing is tricky and has several pitfalls: (1) for effective GPU co-processing, the transfer bottleneck between CPU and GPU has to either be reduced or concealed via clever data placement or caching strategies. (2) when integrating GPU co-processing into a realworld Database Management System (DBMS), the challenge arises that DBMS internals --- such as data structures, query processing and optimization --- are traditionally optimized for CPUs. While there is ongoing research on building GPU-aware database systems \cite{Fang:2007:GQC:1247480.1247606}, no unified GPU-aware DBMS architecture has emerged so far.

The authors in \cite{Bre2014} found that GDBMSs should be in-memory column stores, should use the block-at-a-time processing model and exploit all available processing devices for query processing by using a GPU-aware query optimizer. Thus, main memory DBMSs are similar to GPU-accelerated DBMSs, and most in-memory, column-oriented DBMSs can be extended to efficiently support co-processing on GPUs.

\section{Preliminary Considerations}

The graphics card --- henceforth also called the device --- is connected to the host system via the PCIExpress bus. All data transfer between host and device has to pass through this comparably low-bandwidth bus.

The graphics card itself contains one or more GPUs and a few gigabytes of device memory. Typically, host and device do not share the same address space, meaning that neither the GPU can directly access the main memory nor the CPU can directly access the device memory.

The GPU itself consists of a few multiprocessors, which can be seen as very wide SIMD processing elements. Each multiprocessor packages several scalar processors with a few kilobytes of high-bandwidth, on-chip shared memory, cache, and an interface to the device memory.

Programs that run on a graphics card are written in the so-called kernel programming model. Programs in this model consist of host code and kernels. The host code manages the graphics card, initializing data transfer and scheduling program execution on the device. A kernel is a simplistic program that forms the basic unit of parallelism in the kernel programming model. Kernels are scheduled concurrently on several scalar processors in a SIMD fashion: Each kernel invocation --- henceforth called thread --- executes the same code on its own share of the input. All threads that run on the same multiprocessor are logically grouped into a workgroup.

One of the most important performance factors in GPU programming is to avoid data transfers between host and device: All data has to pass across the PCIexpress bus, which is the bottleneck of the architecture. Data transfer to the device might therefore consume all time savings from running a problem on the GPU. This becomes especially evident for I/O-bound algorithms: Since accessing the main memory is roughly two to three times faster than sending data across the PCIexpress bus, the CPU will usually have finished execution before the data has even arrived on the device.

Graphics cards achieve high performance through massive parallelism. A problem should be easy to parallelize to gain most from running on the GPU. Another performance pitfall in GPU programming is caused by divergent code paths. Since each multiprocessor only has a single instruction decoder, all scalar processors execute the same instruction at a time. If some threads in a workgroup diverge, for example due to data-dependent conditionals, the multiprocessor has to serialize the code paths, leading to performance losses. While this problem has been somewhat alleviated in the latest generation of graphics cards, it is still recommended to avoid complex control structures in kernels where possible.

Currently, two major frameworks are used for programming GPUs to accelerate database systems, namely the Compute Unified Device Architecture (CUDA) and the Open Compute Language (OpenCL). Both frameworks implement the kernel programming model and provide API's that allow the host CPU to manage computations on the GPU and data transfers between CPU and GPU. In contrast to CUDA, which supports NVIDIA GPUs only, OpenCL can run on a wide variety of devices from multiple vendors \cite{gaster2012heterogeneous}. However, CUDA offers advanced features such as allocation of device memory inside a running kernel or Uniform Virtual Addressing (UVA), a technique where CPUs and GPUs share the same virtual address space and the CUDA driver transfers data between CPU and GPU transparently to the application \cite{NVIDIA}.

\section{Non-Functional and Functional Properties of a GPU-aware DBMS (GDBMS) Architecture}
\subsection{Non-Functional Properties -- for which DBMSs are typically optimized for  performance and portability} 

Tsirogiannis and others found that in most cases, the configuration performing best is also the most energy efficient configuration due to the large up-front power consumption in modern servers \cite{Tsirogiannis:2010:AEE:1807167.1807194}.

He and others observed that joins are 2--7 times faster on the GPU, whereas selections are 2--4 times slower, due to the required data transfers \cite{He:2008:RJG:1376616.1376670}. One major point for achieving good performance in a GDBMS is therefore to avoid data transfers where possible.

While the GPU is well suited for easily parallelizable operations
(e.g., predicate evaluation, arithmetic operations), the CPU is the vastly better  fit when it comes to operations that require complex control structures or significant inter-thread communications (e.g., hash table creation or complex user-defined functions). A complex decision model, that incorporates these four factors, is needed to decide on an optimal operator placement \cite{Bre2014}.

Modern DBMSs are tailored towards CPUs and apply traditional compiler techniques to achieve portability across the different CPU architectures (e.g., x86, ARM, Power). By using GPUs --- or generally, heterogeneous coprocessors --- this picture changes, as CPU code cannot be automatically ported to run efficiently on a GPU. Also, certain GPU toolkits --- such as CUDA --- bind the DBMS vendor to a certain GPU manufacturer. In order to achieve optimal performance, each device typically needs its own optimized version of the database operators \cite{broneske2014toward}. We need to take special care to achieve comparable applicability with respect to traditional DBMSs.

\subsection{Functional Properties}
\begin{enumerate}
  \item He and others demonstrated that GPU-acceleration cannot achieve significant speedups if the data has to be fetched from disk, because of the IO bottleneck, which dominates execution costs \cite{He:2009:RQC:1620585.1620588}. Hence, a GPU-aware database architecture should make heavy use of in-memory technology.
  \item A GPU-aware DBMS should use columnar storage. Ghodsnia concluded that a column store is more suitable than a row store, because a column store (1) allows for coalesced memory access on the GPU, (2) achieves higher compression rates (an important property considering the current memory limitations of GPUs), and (3) reduces the volume of data that needs to be transferred. For example, in case of a column store, only those columns needed for data processing have to be transferred between processing devices. In contrast, in a row-store, either the full relation has to be transferred or a projection has to reduce the relation to the data needed to process a query. Both approaches are more expensive than storing the data column wise.
  \item Tuple-wise processing is not possible on the GPU, due to lacking support for inter-kernel communication. A GDBMS should utilize an operator-at-a-time model.
  \item A GDBMS should make use of all available storage and not constrain itself to GPU RAM. While this complicates data processing, and requires a data-placement strategy, it's possible to expect the hybrid to be faster than a pure CPU- or GPU-resident system.
  \item Effective GPU Buffer Management. It is more efficient to transfer few large data sets than many little datasets (with the same total data volume), it could be more beneficial to cache and manage whole columns.
  \item Query Placement and Optimization. Given that a GPU-aware DBMS has to manage multiple processing devices, a major problem is to automatically decide which parts of the query should be executed on which device. Traditional approaches for a distributed system do not take into account specifics of hybrid CPU/GPU systems. Therefore, tailor-made co-processing approaches are likely to outperform approaches from distributed or federated query-processing.
  \item Consistency and Transaction Processing. Keeping data consistent in a distributed database is a widely studied problem. But, research on transaction management on the GPU is almost non-existent.
\end{enumerate}

\textbf{In Summary:} A GPU-aware database system should reside in-memory and use columnar storage. As processing model, it should implement operator-at-a-time bulk processing model, potentially enhanced by dynamic code compilation. The system should make use of all available (co-)processors in the system (including the CPU!) by having a locality-aware query optimizer, which distributes the workload across all available processing resources. In case the GPU-aware DBMS needs transaction support, it should use an optimistic transaction protocol, such as the timestamp protocol. Finally, in order to reduce implementation overhead, the ideal GDBMS would be hardware-oblivious, meaning all hardware-specific adaption is handled transparently by the system itself.

\section{GPU-accelerated DBMS}
The following existing DBMS systems use GPU acceleration. 
\begin{enumerate}
  \item CoGaDB (Universitat Magdeburg, 2013, Open Source)
  \item GPUDB (Ohio State University, 2013, Open Source)
  \item GPUQP (Hong Kong University of Science and Technology, 2007, Open Source)
  \item GPUTx (Nanyang Technological University, 2011)
  \item MapD (Massachusetts Institute of Technology, 2013)
  \item Ocelot (Technische Universitat Berlin, 2013, Open Source)
  \item OmniDB (Nanyang Technological University, 2013, Open Source)
  \item Virginian, (NEC Laboratories America, 2012, Open Source)
\end{enumerate}

 For all eight systems are designed with main-memory databases in mind, keeping a large fraction of the database in the CPU's main memory. GPUQP and MapD also support diskbased data. However, since fetching data from disk is very expensive compared to transferring data over the PCIe bus, MapD and GPUQP also keep as much data as possible in main memory. Therefore, we mark all systems as main memory storage and GPUQP and MapD additionally as disk-based storage.

 All systems store their data in a columnar layout, there is no system using row-oriented storage. One exception is Virginian.

 Most systems support operator-at-a-time bulk processing.

The first group performs nearly all data processing on one processing device (GPUDB, GPUTx, Ocelot, Virginian), whereas the second group is capable of splitting the workload in parts, which are then processed in parallel on the CPU and the GPU (CoGaDB, GPUQP, MapD, OmniDB).

Apart from GPUTx, none of the surveyed GDBMSs support transactions.

 The only GDBMSs having a portable, hardware-oblivious database architecture are Ocelot and OmniDB.

\subsection{Storage Model}

All systems store their data in a columnar layout, there is no system using row-oriented storage. One exception is Virginian, which  stores data mainly column-oriented, but also keeps complete rows inside a table data structure. This representation is similar to PAX, which stores rows on one page, but stores all records column-wise inside a page \cite{Ailamaki:2001:WRC:645927.672367}.

\begin{table}[H]
\center
  \begin{tabular}{|l||c|c||c|c|}
   \hline
     & \multicolumn{2}{c||}{Storage System} & \multicolumn{2}{c|}{Storage Model}\\
    DBMS & Main-Memory Storage & Disk-based Storage & Column Store & Row Store\\
    \hline
    CoGaDB &\checkmark& x &\checkmark& x \\
    GPUDB &\checkmark& x &\checkmark& x \\
    GPUQP &\checkmark&\checkmark&\checkmark& x \\
    GPUTx &\checkmark& x &\checkmark& x \\
    MapD &\checkmark&\checkmark&\checkmark& x \\
    Ocelot &\checkmark& x &\checkmark& x \\
    OmniDB &\checkmark& x &\checkmark& x \\
    Virginian &\checkmark& x &\checkmark& x \\
     \hline
  \end{tabular}
  \caption{Classification of Storage System and Storage Model --- Legend:\checkmark--- Supported, x --- Not Supported, o --- Not Applicable}
\end{table}

\subsection{Processing Model}

The processing model varies between the surveyed systems. The first observation is that no system uses a traditional tuple-at-a-time volcano model. Most systems support operator-at-a-time bulk processing. The only exception is GPUTx, which does not support OLAP workloads, because it is an optimized OLTP engine.

\begin{table}[H]
\center
  \begin{tabular}{|l||c|c|c|}
   \hline
     & \multicolumn{3}{c|}{Processing Model}\\
    DBMS & Operator-at-a-Time & Block-at-a-Time & Just-in-Time Compilation\\
    \hline
    CoGaDB &\checkmark& x & x  \\
    GPUDB &\checkmark&\checkmark&\checkmark \\
    GPUQP &\checkmark& x & x \\
    GPUTx & o & o & o  \\
    MapD &\checkmark&\checkmark&\checkmark \\
    Ocelot &\checkmark& x & x  \\
    OmniDB &\checkmark&\checkmark& x  \\
    Virginian &\checkmark&\checkmark& \checkmark \\
     \hline
  \end{tabular}
  \caption{Classification of Processing Model --- Legend:\checkmark--- Supported, x --- Not Supported, o --- Not Applicable}
\end{table}

\subsection{Query Placement and Optimization}
Two major groups of systems: (1) performs nearly all data processing on one processing device (GPUDB, GPUTx, Ocelot, Virginian), whereas group (2) is capable of splitting the workload in parts, which are then processed in parallel on the CPU and the GPU (CoGaDB, GPUQP, MapD, OmniDB). The systems in the first group are that support only single-device processing (SDP), whereas systems of the second group are capable of using multiple devices and thereby allowing cross-device processing (CDP). The hybrid query optimization approaches of CoGaDB, GPUQP, MapD, and OmniDB are mostly greedy strategies or other simple heuristics. So far, there are no query optimization approaches for machines having multiple GPUs.

\begin{table}[H]
\center
  \begin{tabular}{|l||c|c|c|}
   \hline
     & \multicolumn{2}{c|}{Query Processing}\\
    DBMS & Single-Device Processing & Cross-Device Processing\\
    \hline
    CoGaDB &\checkmark& x\\
    GPUDB &\checkmark&\checkmark \\
    GPUQP &\checkmark& x  \\
    GPUTx & o & o  \\
    MapD &\checkmark&\checkmark \\
    Ocelot &\checkmark& x  \\
    OmniDB &\checkmark&\checkmark \\
    Virginian &\checkmark&\checkmark \\
     \hline
  \end{tabular}
  \caption{Classification of Query Processing --- Legend:\checkmark--- Supported, x --- Not Supported, o --- Not Applicable}
\end{table}

\subsection{Portability}
The only GDBMSs having a portable, hardware-oblivious database architecture are Ocelot and OmniDB. All other systems are either tailored to a vendor specific programming framework or have no technique to hide the details of the device-specific operators in the architecture. Ocelot's approach has the advantage that only a single set of parallel database operators has to be implemented, which can then be mapped to all processing devices supporting OpenCL (e.g., CPUs, GPUs, or Xeon Phis). By contrast, OmniDB uses an adapter interface, in which each adapter provides a set of operators and cost functions for a certain processing-device type. The trend towards hardware-oblivious DBMSs is likely to continue.

\begin{table}[H]
\center
  \begin{tabular}{|l||c||c|c|}
   \hline
     & & \multicolumn{2}{c|}{Portability}\\
    DBMS & Transaction Support & Hardware Aware & Hardware Oblivious \\
    \hline
    CoGaDB & x & \checkmark & x  \\
    GPUDB & x &\checkmark & x  \\
    GPUQP & x & \checkmark & x \\
    GPUTx & \checkmark & \checkmark & x  \\
    MapD & x &\checkmark& x  \\
    Ocelot & x & x & \checkmark  \\
    OmniDB & x & x & \checkmark  \\
    Virginian & x &\checkmark& x \\
     \hline
  \end{tabular}
  \caption{Classification of Transaction Support and Portability --- Legend:\checkmark--- Supported, x --- Not Supported, o --- Not Applicable}
\end{table}

\textbf{In summary,} most main-memory DBMSs supporting column-oriented data layout and bulk processing to be GPU-accelerated DBMSs. The following extension points can be identified: Cost models, CPU/GPU scheduler, hybrid query optimizer, access structures and algorithms for the GPU, and a data placement strategy. Implementing these extensions is a necessary precondition for a DBMS to support GPU co-processing efficiently.

\section{Open Challenges and Research Questions}

Two major classes of challenges: The IO bottleneck, which includes disk IO as well as data transfers between CPU and GPU, and query optimization.

\section{Conclusion}

The future machines will likely consist of a set of heterogeneous processors, having CPUs and specialized co-processors such as GPUs, Multiple Integrated Cores (MICs), or FPGAs. Hence, the question of using co-processors in databases is not why but how we can do this most efficiently.

The pioneer of modern co-processors is the GPU, and many prototypes of GPU-accelerated DBMSs have emerged over the past years implementing new co-processing approaches and proposing new system architectures. A GDBMS should be an in-memory, column-oriented DBMS using the block-at-a-time processing model, possibly extended by a just-in-time-compilation component. The system should have a query optimizer that is aware of co-processors and data-locality, and is able to distribute a workload across all available (co-)processors. The results are not limited to GPUs, but should also be applicable to other co-processors. The existing techniques can be applied to virtually all massively parallel processors having dedicated high-bandwidth memory with limited storage capacity.

\bibliography{HUAWEI_Accelerator} 
\bibliographystyle{ieeetr}
\end{document}